\documentclass[twocolumn,pre,superscriptaddress]{revtex4}

\usepackage{graphicx}
\usepackage{amsmath}

\newcommand\etal{\textit{et~al.}}

\begin{document}

\title{Coauthorship and citation in scientific publishing}

\author{Travis Martin}
\affiliation{Department of Electrical Engineering and Computer Science,
  University of Michigan, Ann Arbor, MI 48109, U.S.A.}
\author{Brian Ball}
\affiliation{Department of Physics, University of Michigan, Ann Arbor,
MI 48109, U.S.A.}
\author{Brian Karrer}
\affiliation{Department of Physics, University of Michigan, Ann Arbor,
MI 48109, U.S.A.}
\author{M. E. J. Newman}
\affiliation{Department of Physics, University of Michigan, Ann Arbor,
MI 48109, U.S.A.}
\affiliation{Center for the Study of Complex Systems, University of
  Michigan, Ann Arbor, MI 48109, U.S.A.}

\begin{abstract}
  A large number of published studies have examined the properties of
  either networks of citation among scientific papers or networks of
  coauthorship among scientists.  Here, using an extensive data set
  covering more than a century of physics papers published in the Physical
  Review, we study a hybrid coauthorship/citation network that combines the
  two, which we analyze to gain insight into the correlations and
  interactions between authorship and citation.  Among other things, we
  investigate the extent to which individuals tend to cite themselves or
  their collaborators more than others, the extent to which they cite
  themselves or their collaborators more quickly after publication, and the
  extent to which they tend to return the favor of a citation from another
  scientist.
\end{abstract}

\maketitle

\section{Introduction}

Citation networks~\cite{Price65} and coauthorship
networks~\cite{Grossman95,Grossman02,Newman01} are distinct network
representations of bodies of academic literature that have both been the
subject of quantitative analysis in recent years.  In a citation network
the network nodes are papers and a directed edge runs from paper~A to
paper~B if A cites B in its bibliography.  In a coauthorship network the
nodes are authors and an undirected edge connects two authors if they have
written a paper together.  Both kinds of network can shed light on habits
and patterns of academic research.  Citation networks, for instance, can
give a picture of the topical connections between papers, while
coauthorship networks can shed light on patterns of collaboration such as
the size of collaborative groups or the frequency of repeated
collaboration.

However, there are also many interesting questions that can be answered
only by combining citation and coauthorship data, questions such as how
much researchers cite their collaborators relative to others in their
field, or whether a researcher is more likely to cite others from whom they
previously received a citation.  There has been relatively little work on
questions like these to date~\cite{Milojevic12}, and it is on these
questions that we focus in this paper.

We analyze a large data set made available by the American Physical Society
(APS), which consists of bibliographic and citation data for the Physical
Review~\footnote{More details about the data set can be found on the web at
  https://publish.aps.org/datasets}.  The Physical Review is a family of
journals operated by the APS and published continuously for over a century
with articles covering all aspects of physics.  The data set we analyze
runs from the journals' inception in 1893 to 2009 and describes nearly half
a million papers, including their authorship and the citations between
them.  A number of previous analyses of these data have been
published~\cite{Redner05,Chen10,Gualdi11}, but our work adopts a somewhat
different viewpoint from other studies in focusing on the interactions
between authorship and citation.  Among other things, we find, for example,
that researchers cite their own or coauthors' papers more quickly after
publication than they do the work of others; that authors show a strong
tendency to return the favor of a citation from another author, especially
a previous coauthor; that, contrary to some recent conjectures, having a
common coauthor does not make two authors likely to collaborate in
future~\cite{Newman01a, Davidsen02, Holme02}; and that there has not (at
least within the journals we study) been any increase over time in
self-citations, the number holding roughly constant at about 20\% of all
citations for over a century.

\section{The data set}
\label{sec:data}
In its raw form the data set we study contains records for $462\,090$
papers published in the various Physical Review journals, each identified
with a unique numerical label.  Data for each paper include paper title,
date of publication, the published names and affiliations of each of the
authors, and a list of the numerical labels of previous Physical Review
papers cited.  The data set is unusual in two respects: the long period of
time it covers, which spans 116 years from 1893 to 2009, and the fact that
it includes citation data and hence allows us to compare coauthorship
patterns with citations, at least for that portion of the citation network
that appears in the Physical Review---citations to and from
non-Physical-Review journals, of which there are many, are not included.

Before performing any analysis, however, there are some hurdles to
overcome.  Foremost among them is the fact that the name of an author alone
does not necessarily identify him or her uniquely.  Two authors may have
the same name, or the same author may be identified differently in
different publications (with or without a middle initial, for example).
Unlike some journals, such as those of the American Mathematical
Society~\footnote{A description of the unique author identifier system used
  by the American Mathematical Society can be found at
  http://www.istl.org/01-summer/databases.html.}, the Physical Review does
not maintain unique author identifiers that can be used to attribute
authorship unambiguously.  As a first step in analyzing the data,
therefore, we have processed it using a number of disambiguation techniques
in order to infer actual author identity from author names as accurately as
possible.  Details of the disambiguation process are given in
Appendix~\ref{app:data_processing}.

In addition, we have performed a modest culling of the data to remove
outliers, the most substantial action being the removal of all papers with
fifty or more authors, which are primarily recent papers in experimental
high-energy physics.  (Almost all of them, about 91\%, were published
either in Physical Review D, which covers high-energy physics, or Physical
Review Letters; the remainder were in Physical Review C, which covers
nuclear physics.)  As we show shortly, though papers with more than fifty
authors are only a small fraction of the whole (about 0.7\%), their
inclusion skews results for the last thirty years substantially by
comparison with the rest of the time period.  For results whose outcome
depends strongly on the presence or not of these papers, we quote results
both with and without, for comparison.

\begin{table}
\begin{center}
\begin{tabular}{lcc}
 & & Papers with 50 \\ & All papers & authors or fewer \\ \hline
Total papers & 460889 & 457516 \\
Total authors & 235533 & 226641 \\
Authors per paper & 5.35 & 3.34 \\
Citations per paper & 10.16 & 10.16 \\
Number of collaborators & 59.44 & 17.24 \\
Papers per author & 10.47 & 6.74
\end{tabular}
\end{center}
\caption{Mean values of some statistics for our data set, with and without
  papers having over 50 authors.}
\label{tab:summary}
\end{table}

Table~\ref{tab:summary} gives some basic parameters of the resulting data
set.

\section{Analysis}
In the next few sections we present a variety of analyses of the Physical
Review data set.  We begin by looking at some basic parameters of
authorship and coauthorship.

\begin{figure}
  \includegraphics[width=\columnwidth,clip=true]{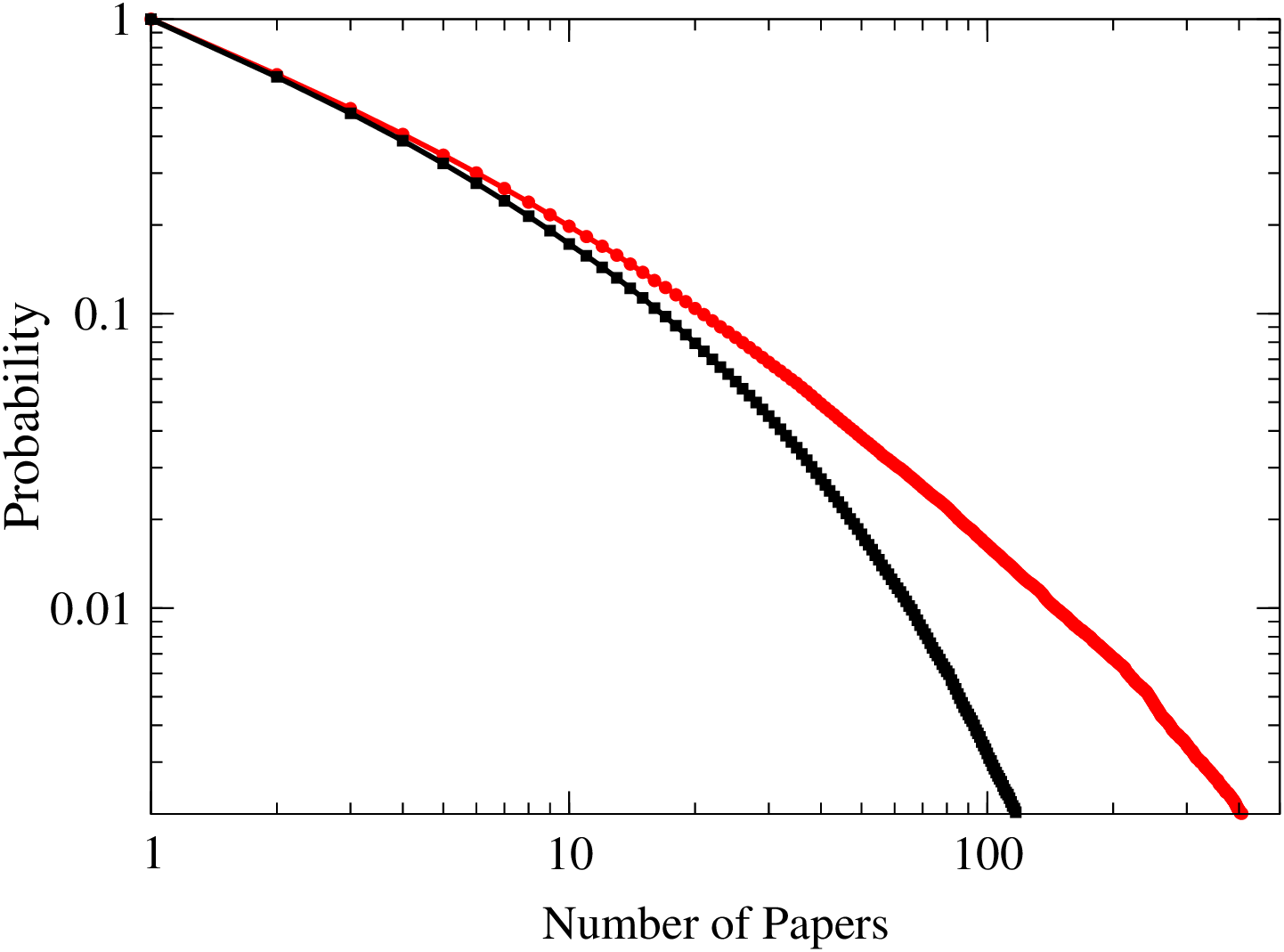}
  \caption{Probability that an author wrote more than a given number of
    papers.  Red circles indicate values calculated from the full data set;
    black squares are values from the data set after papers with fifty or
    more authors have been removed.  The plot is cut off around 500 papers
    because there is very little data beyond this point.}
  \label{fig:papers_published}
\end{figure}

\subsection{Authorship patterns}
Figure~\ref{fig:papers_published} shows a cumulative distribution function
for the number of papers an author publishes, aggregated over the entire
data set.  That is, the figure shows the fraction of authors who published
$n$ papers or more as a function of~$n$, which is a crude measure of
scientific productivity.  The axes in the figure are logarithmic, and the
approximate straight-line form of the distribution function implies that
scientific productivity follows, roughly speaking, a power law, a result
known as Lotka's law, first observed by Alfred Lotka in 1926~\cite{Lotka26}
and confirmed by numerous others since.  (It has also been suggested that
the distribution is log-normal rather than power-law~\cite{Shockley57}.  It
is known to be hard to distinguish empirically between log-normal and
power-law distributions~\cite{CSN09}.)  In Fig.~\ref{fig:papers_published}
we give separate curves with and without the papers that have fifty or more
coauthors.  As the figure shows, the difference between the two is
primarily in the tail of the distribution, among the authors who have
published the largest number of papers, indicating that a significant
fraction of the most productive authors are those in large collaborations.
In fact, if one compiles a list of the fifty authors publishing the largest
numbers of papers, only one of them remains on that list after papers with
fifty or more authors are excluded.  This probably results from a
combination of two effects: first, larger groups can publish more papers
simply because they have more people available to write them; and second, a
large and productive group of collaborators contributes many apparently
prolific authors to the statistics---each of the many coauthors separately
gets credit for being highly productive.  It is precisely because of biases
of this kind that we exclude papers with many authors from some of our
calculations.

\begin{figure}
  \includegraphics[width=0.85\columnwidth,clip=true]{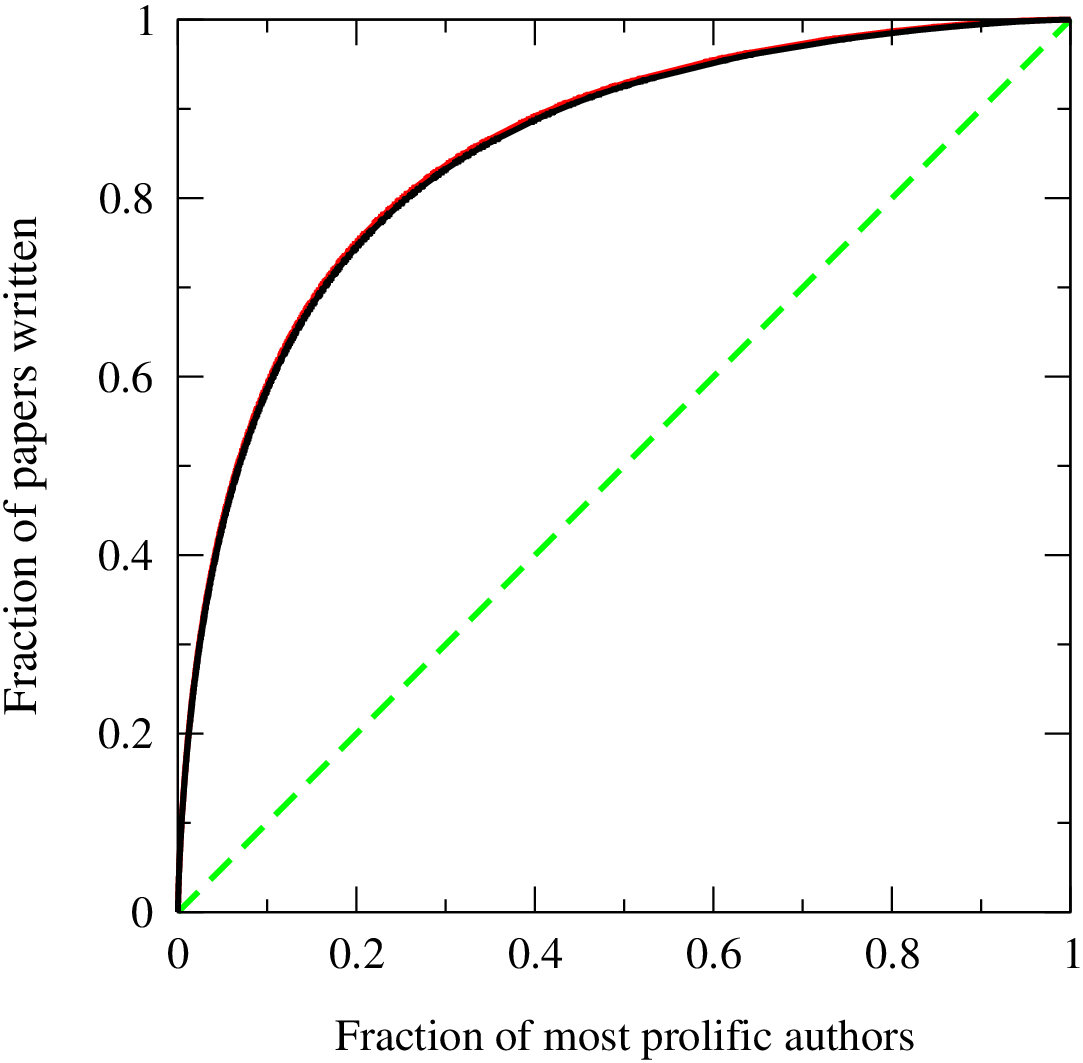}
  \caption{Fraction of papers written by the most prolific authors (with
    credit for multi-author papers divided among coauthors, as described in
    the text).  The red curve represents values calculated from the full
    data set; the black curve represents values after papers with fifty or
    more authors have been removed.  Note that the two curves are almost
    indistinguishable.  The dashed line indicates the form the curve would
    take if all authors published the same number of papers.}
  \label{fig:gini}
\end{figure}

We can remedy this problem to some extent by measuring productivity in a
more sophisticated fashion.  Rather than just counting up all the papers an
author was listed on, we can instead divide up the authorship credit for a
paper among the contributing authors so that, for example, each author on a
two-author paper is credited with half an authorship for that paper.  This
reduces significantly the impact of large collaborations on the statistics,
though the distribution of number of papers authored is still highly
skewed, with certain authors producing much more science than others.  A
common way to visualize such skewed distributions is to use a Lorenz curve,
a plot of the fraction of papers produced by the most prolific authors
against the fraction of authors that produced them.  Such a curve is shown
for our data set in Fig.~\ref{fig:gini}, and the sharp rise in the curve at
the left-hand side indicates the concentration of scientific productivity
among the most productive scientists.  Note for instance that productivity
appears roughly to follow the so-called 80--20 rule, such that about 80\%
of the output is produced by the 20\% most productive authors.  Notice also
that there is almost no difference in the Lorenz curves with and without
the 50-plus-author papers, precisely because we have divided up the
authorship credit so that the effect of many-author papers is diminished.

The distribution can be further quantified by measuring a Gini coefficient,
which is defined as the excess area under the Lorenz curve compared to the
case where everyone has the exact same productivity.  In our data set, the
Gini coefficient is $0.70$, a relatively large figure as such coefficients
go, indicating high skew.  (Gini coefficients for wealth inequality, for
example, which is the context in which such coefficients are perhaps best
known, rarely rise above 0.6, even in the most inequitable countries.)

\begin{figure}
  \includegraphics[width=\columnwidth,clip=true]{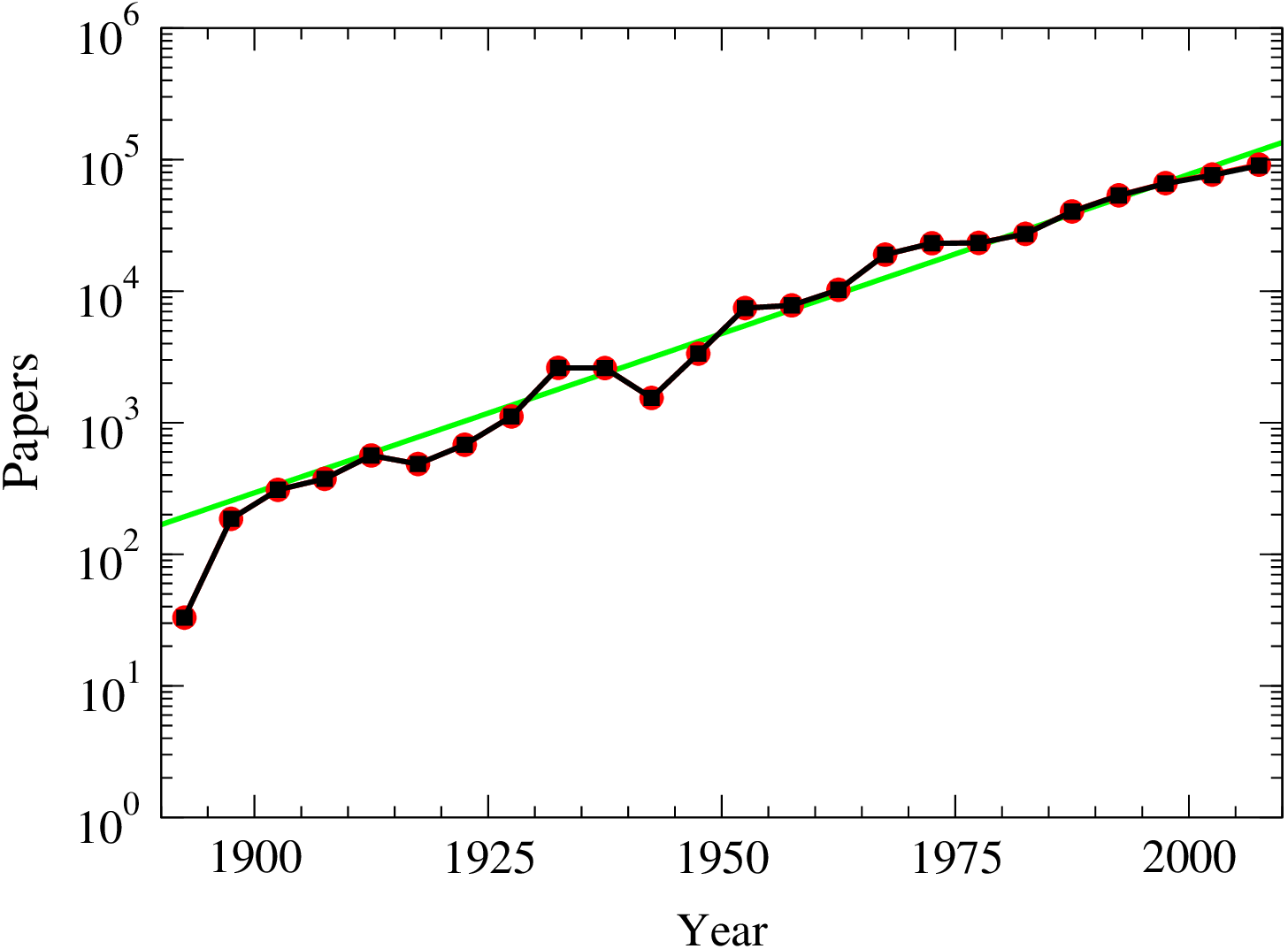}
  \caption{Number of papers published in each five-year block.  Red circles
    indicate numbers calculated from the full data set; black squares are
    calculated from the data set after papers with fifty or more authors
    have been removed.  Note that the two values are almost
    indistinguishable.  The straight line is the best-fit exponential.}
  \label{fig:num_papers}
\end{figure}

The data set also allows us to measure the productivity of the entire field
of physics over time, something that cannot be done with many other data
sets.  Figure~\ref{fig:num_papers} shows the total number of papers
published in the Physical Review in five year time blocks since 1893.  With
the important caveat that these results are for a single collection of
journals only, and one moreover whose role within the field has evolved
during its history from provincial up-start to one of the leading physics
publications on the planet, we see that there is a steady increase in the
volume of published work, which appears roughly to follow an exponential
law (a straight line on the semi-logarithmic scales of the figure).  An
interesting feature is the dip in the curve in the 1940s, which coincides
with the second World War, followed by a recovery in the 1950s, perhaps
attributable in part to increased science funding in the postwar period.
The combined result of these deviations, however, is only to put the curve
back on the same path of exponential growth after the war that it was
already on before it.  In his early studies of secular trends in scientific
output, Derek de Solla Price~\cite{Price61,Price63} noted a similar
exponential growth interrupted by the war, and measured the doubling time
of the growth process to be in the range from 10 to 15 years.  The best
exponential fit to our data gives a compatible figure of 11.8 years.

\begin{figure}
  \includegraphics[width=\columnwidth,clip=true]{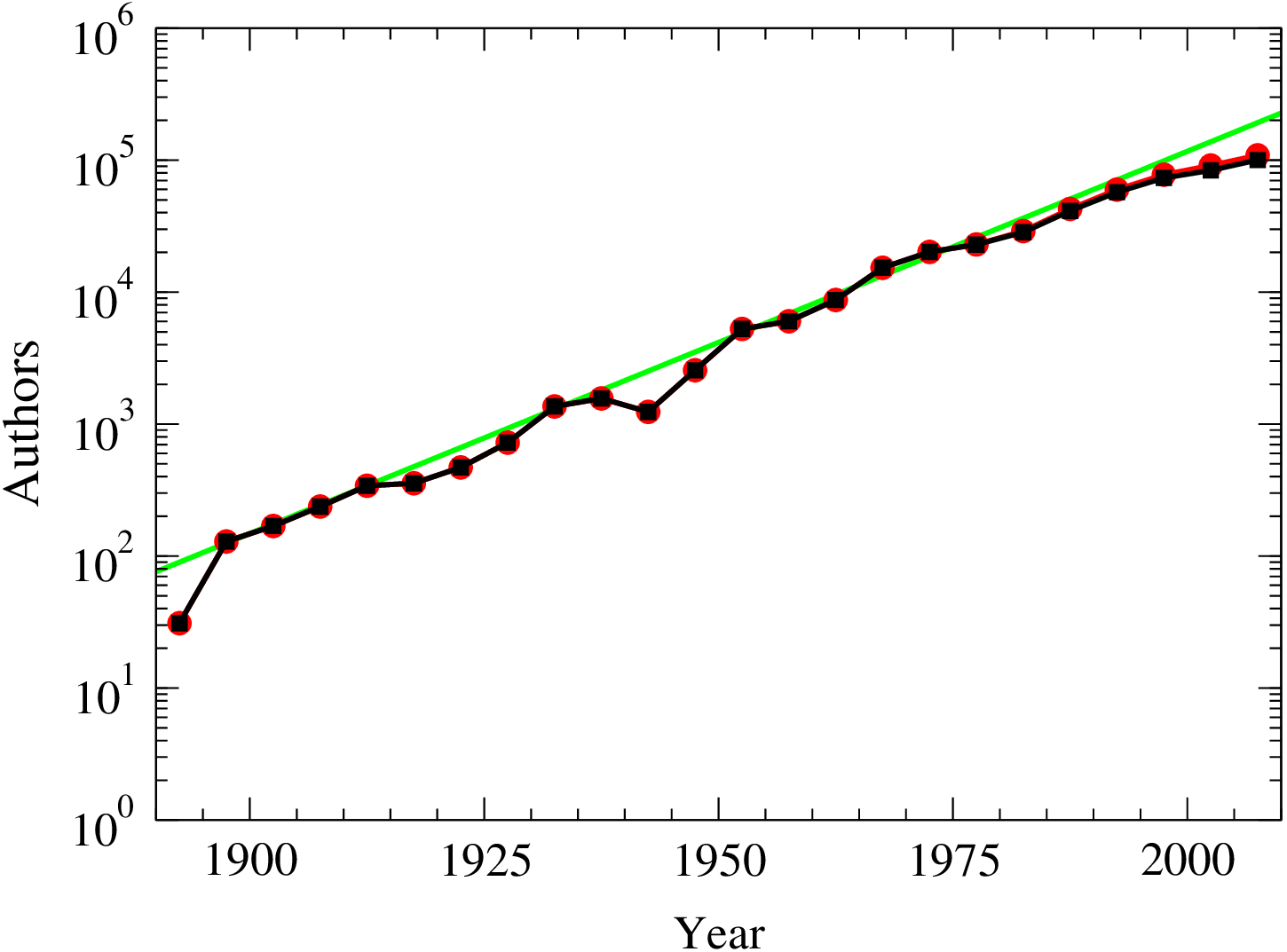}
  \caption{Number of unique authors who published a paper in each five-year
    block.  Red circles indicate numbers calculated from the full data set,
    while black squares are calculated from the data set after papers with
    fifty or more authors have been removed.  Note that the two values are
    almost indistinguishable.  The straight line is the best-fit
    exponential.}
  \label{fig:num_authors}
\end{figure}

Figure~\ref{fig:num_authors} shows the corresponding plot of the number of
unique authors in the data set in each five-year block as a function of
time.  Like the number of papers published, the number of authors appears
to be increasing exponentially, and with a roughly similar (but slightly
smaller) doubling time of 10.4 years.  Thus, despite the marked increase in
productivity of the field as a whole, it appears that each individual
scientist has produced a roughly constant, or even slightly decreasing,
number of papers per year over time.

\begin{figure}
  \includegraphics[width=\columnwidth,clip=true]{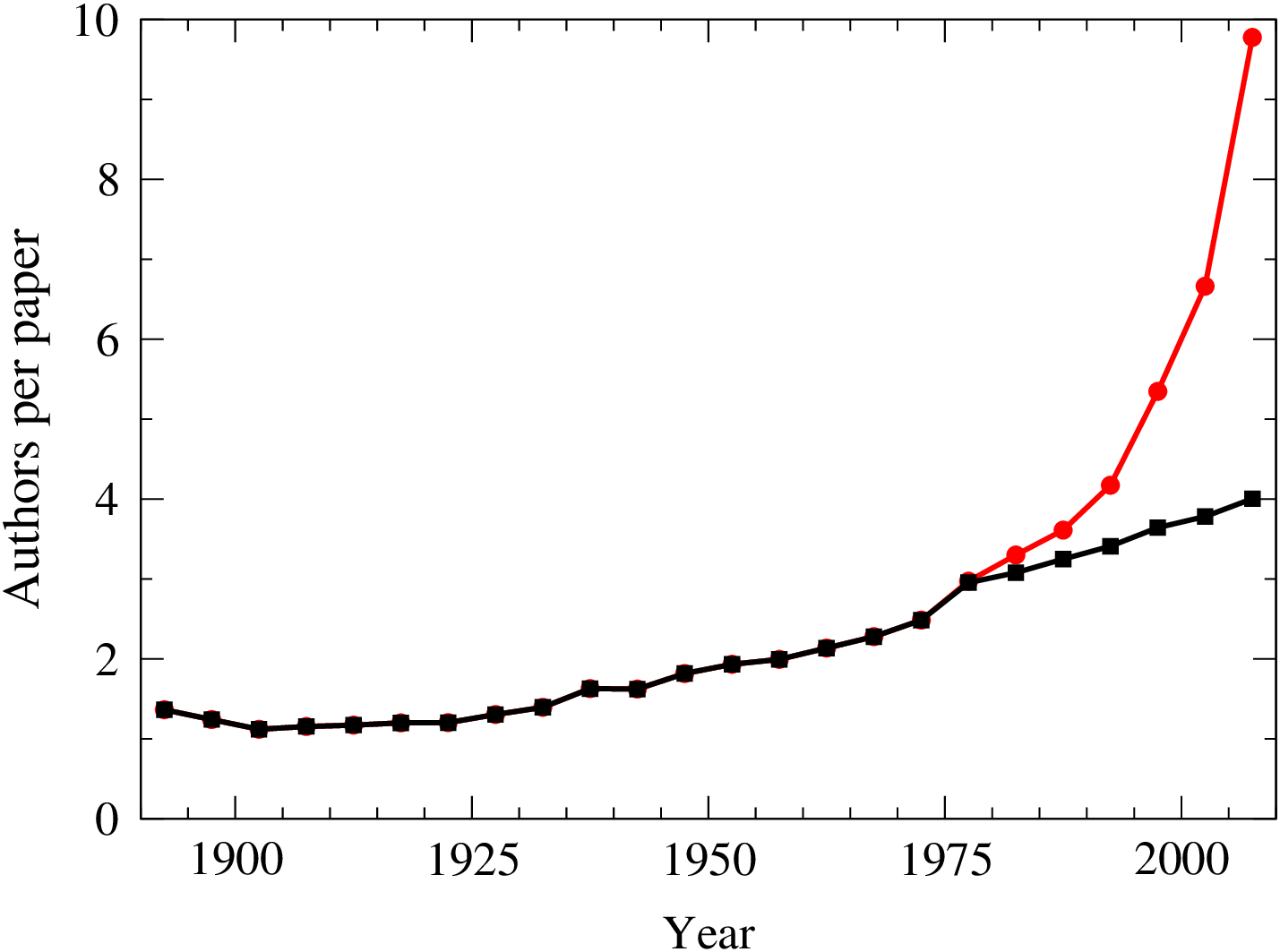}
  \caption{Number of authors per paper averaged over five-year blocks.  Red
    circles indicate the full data set; black squares are the data set
    after papers with fifty or more authors have been removed.}
  \label{fig:paper_auths}
\end{figure}

The natural complement to measurement of the number of papers per author is
measurement of the number of authors per paper, i.e.,~the size of
collaborative groups.  Figure~\ref{fig:paper_auths} shows the mean number
of authors per paper in our data set as a function of time, and there is a
clear increasing trend throughout most of the time period covered, with the
average size of a collaborative group rising from a little over one a
century ago to about four today.  A similar effect has been noted
previously by, for example, Grossman and Ion~\cite{Grossman02}, for the
case of mathematics collaborations.  In our calculations we have again
calculated separate curves with and without papers having fifty or more
authors and a comparison between the two reveals a startling effect: while
there is almost no difference at all between the curves prior to about
1975, there is a large and rapidly growing gap between them in the years
since.  Without these papers the growth in group sizes has been slow and
steady for decades; with them it departs dramatically from historical
trends after the 1970s, indicating a large and growing role in physics (or
at least in physics publication) for big collaborations.

\begin{figure}
  \includegraphics[width=\columnwidth,clip=true]{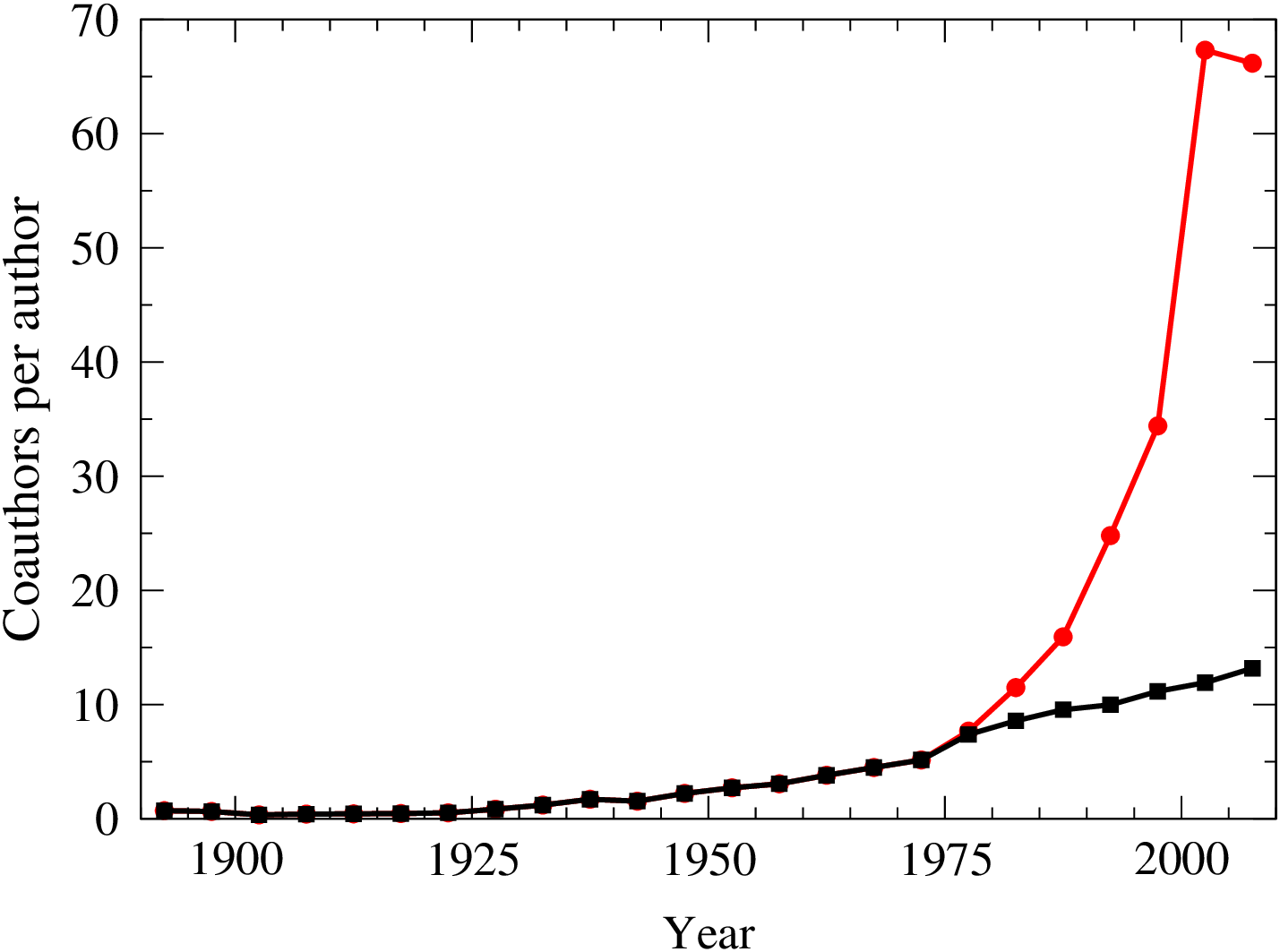}
  \caption{Average number of unique coauthors of an author, averaged in
    five-year blocks.  Red circles indicate the full data set; black
    squares are the data set after papers with fifty or more authors have
    been removed.}
  \label{fig:coauths}
\end{figure}

An alternative view of the same trend is given in Fig.~\ref{fig:coauths},
which shows the number of unique coauthors an author has, on average,
during each five year time block.  Every coauthor in a time block is
counted, even if he or she was also counted in a previous time block (but
previous coauthors are not counted unless they are also coauthors in the
new time block).  As the figure shows, this number has also risen
significantly over the last century, from a little over one to more than
ten today (and more than sixty if one includes collaborations with fifty or
more members).  Since we only have data from the Physical Review, it is
likely that we miss some collaborators, so these numbers are in practice
only lower bounds on the actual numbers.

\subsection{Citation patterns}
\label{sec:cites}

Let us now add the citation portion of the data set to our analyses and
examine citation patterns over time in the Physical Review, as well as
interactions between citation and coauthorship.

\begin{figure}
  \includegraphics[width=\columnwidth,clip=true]{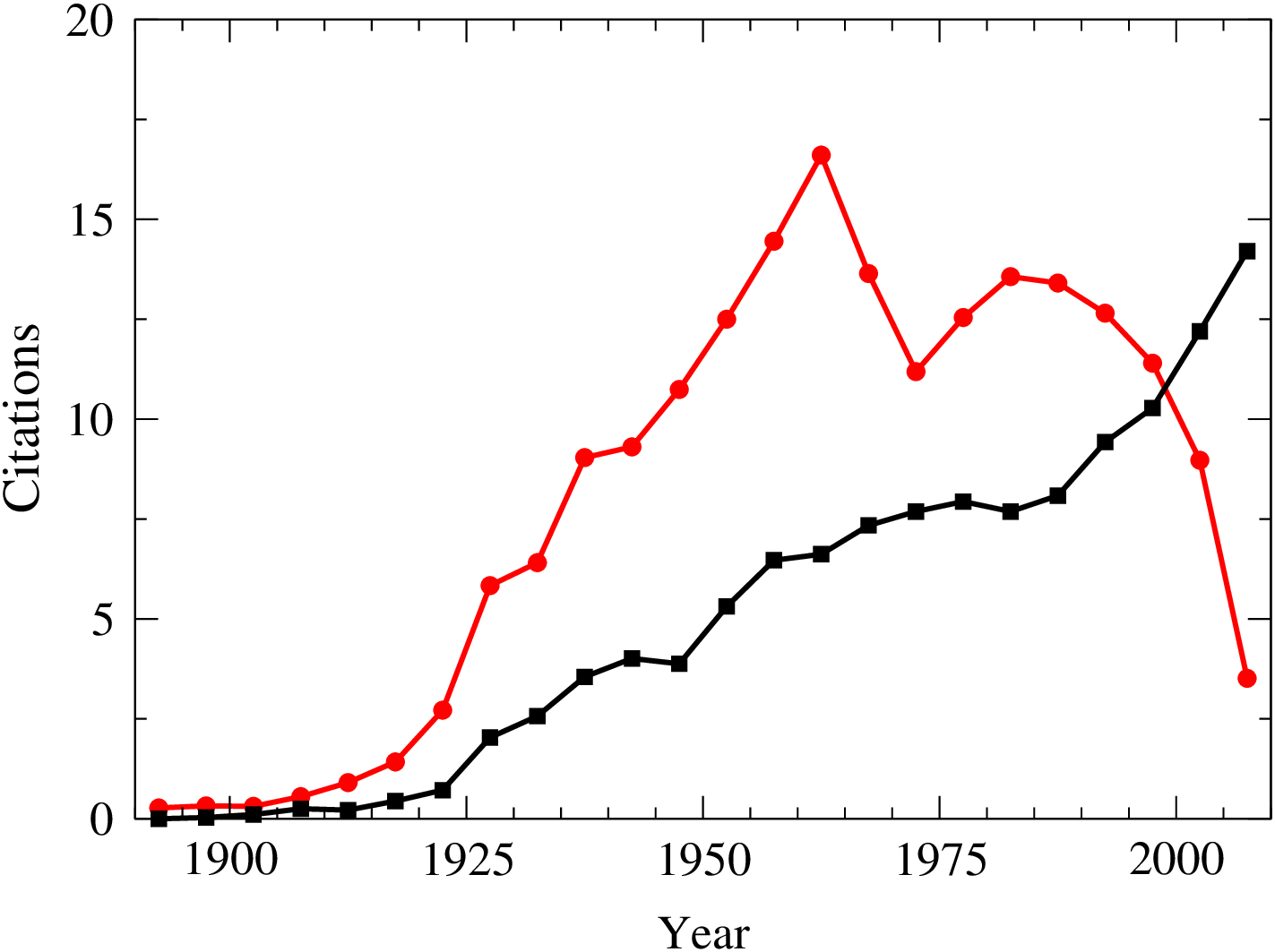}
  \caption{Average numbers of citations made (black squares) and received
    (red circles) per paper, in five-year blocks.}
  \label{fig:cites_time}
\end{figure}	

Figure~\ref{fig:cites_time} shows the average number of citations by a
paper and to a paper, over the time period covered by the Physical Review
data set.  The black curve, the number of citations that a paper makes,
shows a steady increase over time---authors used to cite fewer papers and
have been citing steadily more in recent decades.  One possible explanation
for this phenomenon is the increase in the volume of literature available
to be cited, although it has also been conjectured that authors have been
under greater pressure in recent decades, for example from journal editors
or referees, to add more copious citations to papers~\cite{Wilhite12}.

The red curve in Fig.~\ref{fig:cites_time} is the average number of
citations received by a paper, which shows more irregular behavior, rising
to a peak twice before dropping off in recent times.  A number of effects
are at work here.  First, if (as we will shortly see) most citations are to
papers in the recent past, then a steady increase in citations \emph{by}
papers should lead to an increase in citations \emph{to} papers published
slightly earlier.  Behavior of this kind has been observed in previous
studies, such as the comprehensive study by Wallace~\etal\ using data from
the Web of Science~\cite{Wallace09}.  The growth in number of citations
received cannot continue to the very end of the data set, however, since
the most recent papers are too recent to have accrued a significant number
of citations and hence we expect a drop at the rightmost end of the curve,
as seen in the figure.

There is, however, also a notable dip in the red curve around 1970, whose
origin is less clear.  (It is not seen, for instance, in the work of
Wallace~\etal)\ \ In examining the data for this period in detail, we find
that the dip in citations per paper is due primarily to an increase in the
number of papers published in the Physical Review (which expanded
considerably during this period), while the number of citations received by
those papers, in aggregate, remains roughly constant.  The increase in
papers published may have been in part a response to the general expansion
of US physics research during the 1960s, following the establishment of the
National Science Foundation, but the data indicate that the greater volume
of research did not, at least initially, result in a greater number of
citations received, and hence the ratio of the two displays the dip visible
in Fig.~\ref{fig:cites_time}.  However, the upward trend in the curve
reestablishes itself from about 1970 onward, suggesting that in the long
run there was an increase not only in the number of papers published, but
also in the number that are influential enough to be later cited.

It is interesting to compare the data for citations received with the
predictions of theoretical models for the citation process.  Perhaps the
best known class of models are the preferential attachment
models~\cite{BA99b}, and particularly the 1976 model of
Price~\cite{Price76}, a simple model in which the rate at which a paper
receives citations is assumed to vary linearly with the number it already
has.  In its most naive application, this model makes predictions that
differ strongly from the observations plotted in Fig.~\ref{fig:cites_time}.
The model predicts that the largest number of citations should go to the
oldest papers and the smallest to the youngest, so that the red curve in
the figure should be monotonically decreasing.  There are a number of
possible explanations for the disagreement.  A popular theory is that
papers ``age'' over time, becoming less well cited as they become
older~\cite{Zhu03,Sanyal07}, perhaps because their field has moved on to
other things, because they have been superseded by more advanced or
accurate work, or because their results are so well known that authors no
longer feel the need to cite them.  Were this the case, most citations
would be to recent papers, and the curve of citations received would mostly
mirror the curve of citations given, albeit with a time lag whose length
would be set by the rate at which papers age.  An alternative theory, for
which there is some empirical evidence, is that preferential attachment
models do represent citation patterns quite well within individual
subfields~\cite{Newman09}, but not when applied to the literature as a
whole.  A central parameter in the preferential attachment models is the
date of the start of a subfield, and since different subfields have
different start dates, the model might be expected to work within subfields
but not for the overall data set.

\begin{figure}
  \includegraphics[width=\columnwidth,clip=true]{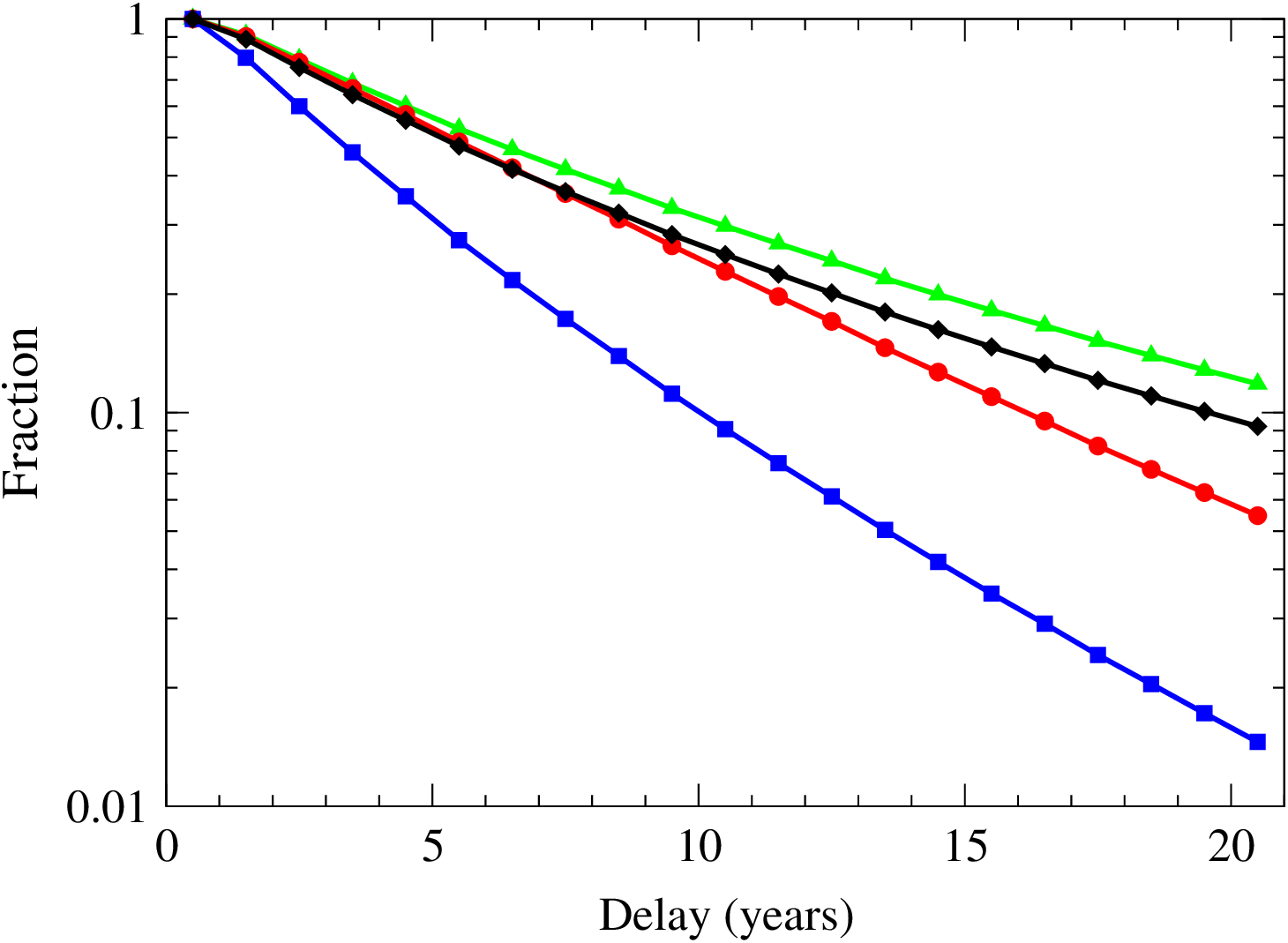}
  \caption{Fraction of citations made more than a given number of years
    after publication.  Black diamonds include all citations, blue squares
    are self-citations, red circles are co-author citations, and green
    triangles are distant citations.}
  \label{fig:cite_delay}
\end{figure}

Figure~\ref{fig:cite_delay} tests the aging of papers within the Physical
Review data set by plotting the fraction of citations that are to papers a
certain time in the past.  Let us focus for the moment on the black curve,
which includes all citations in the entire data set.  The figure shows that
there does indeed appear to be a strong aging effect, with the citation
rate dropping off approximately exponentially over time (which would be a
straight line on the semi-logarithmic scales of the plot).  This finding is
in agreement with previous studies of aging~\cite{Zhu03}, which also found
exponential decay.  An alternative interpretation of the data, however, is
that there is no aging occurring at all, and that the drop in citations is
a purely mechanical effect that results from dilution of the
literature---in a small, young field there are only a few papers to cite
and hence each receives a lot of citations; in an older field there are
more papers and so individual citation rates fall off.  To the extent that
it has been tested, the latter theory appears to agree well with available
citation data and also with the prediction of the preferential attachment
models~\cite{KN09}, so at present the evidence for (or against) aging in
our data set is inconclusive.

\begin{figure}
  \includegraphics[width=\columnwidth,clip=true]{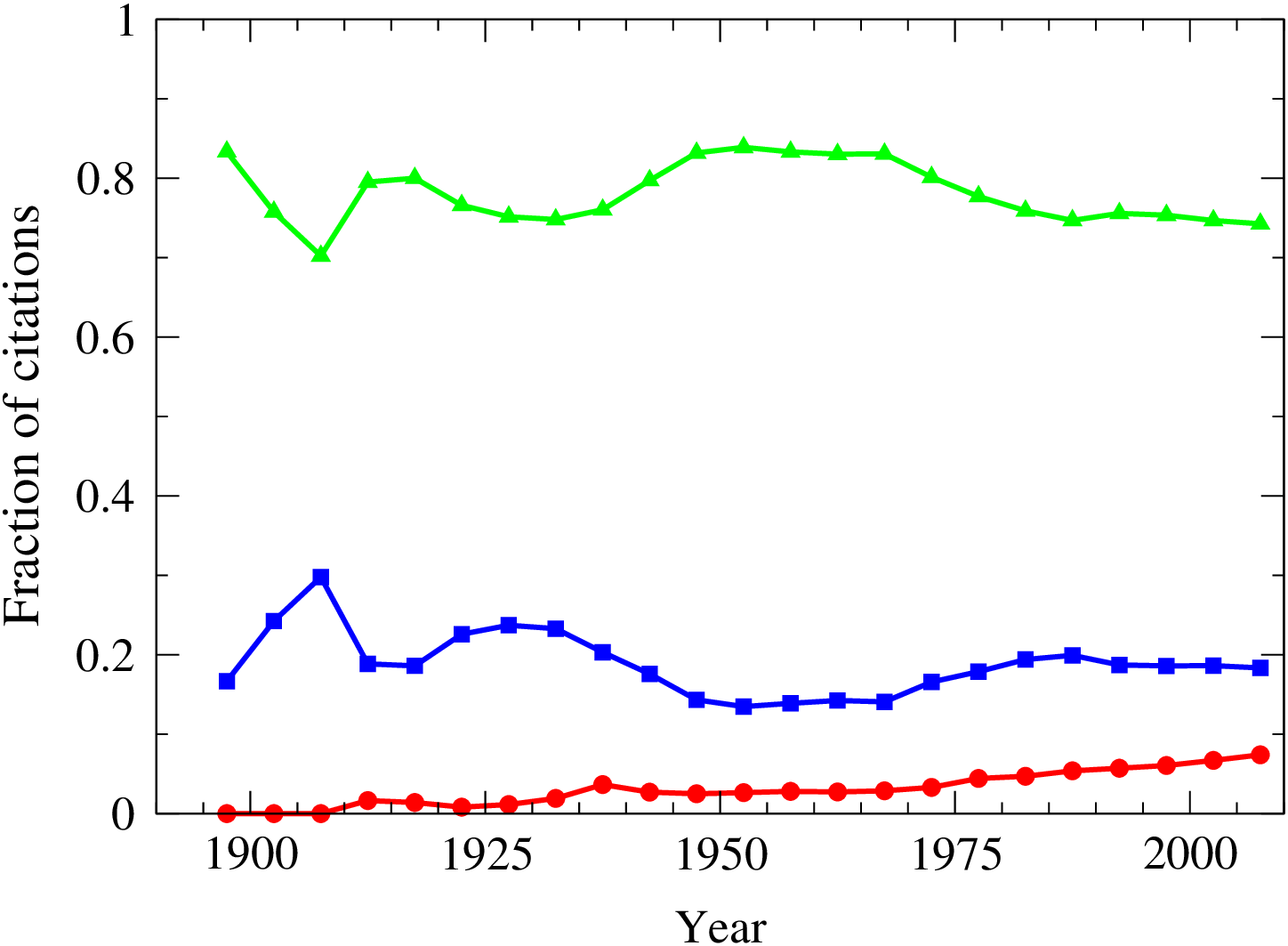}
  \caption{Fraction of citations made, by type, in five-year blocks.  There
    were no citations made in the 1890--1894 block.  Blue squares represent
    self-citations, red circles are co-author citations, and green
    triangles are distant citations.}
  \label{fig:frac_cites}
\end{figure}

\subsection{Interactions between citation and coauthorship}
Perhaps the most interesting aspect of the Physical Review data, however,
is the window it gives us on the interplay between citation and
coauthorship.  One way to probe this interplay is to divide citations
according to the collaborative roles assumed by the authors of the citing
and cited papers and then compare the resulting citation patterns.  In the
present work, we divide citations into three classes, following
Wallace~\etal~\cite{Wallace12}: self-citations, where the citing and cited
papers shared at least one coauthor; coauthor citations, where at least one
author of the citing paper has previously collaborated with at least one
author of the cited paper (but there are no common authors between papers,
so that self-citations and coauthor citations are disjoint); and distant
citations, which includes all citations other than self-citations and
coauthor citations.  (Other authors who have examined citation and
collaboration have gone further and considered also citations between
coauthors of coauthors~\cite{Wallace12}, but this proves computationally
unfeasible in the present case because of the size of the Physical Review
data set.)  We emphasize that we only consider individuals to be coauthors
if they have \emph{previously} coauthored when the citation occurs.
Coauthorship that comes after the citation is not counted.  Also our data
are limited to the Physical Review, so the number of coauthor citations
will in reality be higher than presented here, both because some citations
are missing from our data and because some coauthorships are.

Figure~\ref{fig:frac_cites} shows the fraction of citations that fall into
each of the three classes as a function of the year of publication of the
citing paper. Roughly speaking, the three curves appear flat over time.
There is a modest increase in the fraction of coauthor citations (the
lowest, red curve in the figure), but this can be explained by the increase
in the number of coauthors available for citation, shown in
Fig.~\ref{fig:coauths}, which is of a similar magnitude.  In other
respects, the rule of thumb seems to be that a constant 20\% or so are
self-citations, 75 or 80\% are distant citations, and the small remaining
fraction are to coauthors.

The distribution of time between the publication dates of a new paper and
the papers it cites is shown for the three classes of citation in
Fig.~\ref{fig:cite_delay}, as the blue, red, and green curves.  Here we do
notice a significant difference between the classes.  In particular, the
self-citations (in blue) fall off faster than coauthor and distant
citations.  This implies that a larger fraction of self-citations occur
rapidly after publication, compared with citations in the other classes.
This is not unexpected, given that a researcher presumably knows about
their own research sooner, and in more detail, than they know about
others'.  We note also that coauthor citations are slightly earlier than
distant citations, which again seems reasonable.  One must be careful in
the interpretation of these results, however.  An alternative explanation
for the same observations is that a paper can be cited by others long after
the author retires or leaves the field, which could make the average delay
for citations by others longer than that for self-citation.  There is no
way to tell, purely from the delay statistics themselves, which explanation
is the better one.

\begin{table}
\begin{center}
\begin{tabular}{r|c}
Citation type & Mean delay (years) \\
\hline
Self-citations & 4.12 \\
Coauthor citations & 6.92 \\
Distant citations & 9.02 \\
\hline
All citations & 7.89 \\
\end{tabular}
\end{center}
\caption{Mean time delay between a paper's publication date and the dates
  of the papers it cites.}
\label{tab:delays}
\end{table}

Table~\ref{tab:delays} summarizes the mean delay to citation for the three
citations classes.  We explore the differences between citation classes
further in the next section.

\begin{table}
\begin{center}
\begin{tabular}{r|c|c}
Citation type & Made (\%) & Received (\%) \\ \hline
Self-citation & 68.9 & 60.3 \\
Coauthor citation & 42.0 & 31.3 \\
Both & 35.6 & 26.3 \\
Either & 75.0 & 64.2 \\
Either given both possible & 76.4 & 66.4
\end{tabular}
\end{center}
\caption{Percentage of papers that make or receive at least one citation of a
  given type.}
\label{tab:citation}
\end{table}

\subsection{Self-citation and coauthor citation}
Consider Table~\ref{tab:citation}, which gives the percentages of papers
that make or receive at least one self-citation or coauthor citation,
provided that such a citation is possible.  Nearly 70\% of papers cite at
least one paper by the same author (or one of the same authors, if there
are several), and 60\% of them receive such a citation.  These numbers may
at first appear large, and raise concerns, given the use of citation counts
as a measure of impact, that authors might be inflating their counts by
self-citing~\cite{Hirsch05,Bartneck11}.  But taken with the fact that the
number of citations per paper and the fraction which are self-citations are
both sizable, these large numbers are not unexpected.
Figure~\ref{fig:frac_cites} shows that overall self-citation has remained
constant and moderate, around 20\%, and that there has been no sizable
recent excess in self-citation.

A more interesting question is whether researchers have a tendency to
reciprocate citations by others.  If author A cites a paper of author~B,
does B return the favor by later citing A?  To address this question we
measure the fraction of citations of one author by another (excluding
citations of one's own papers) that are reciprocated in one or more later
publications.  We calculate separate figures for pairs of authors who have
previously co-authored a paper and those who have not and find that 13.5\%
of citations between non-coauthors are reciprocated when possible, while an
impressive 43.8\% of citations between coauthors are reciprocated.  (Keep
in mind that no authors can overlap between a citing and a cited paper for
the citation to be considered a coauthor citation and not a self-citation.)
Both these numbers are very high compared to the expected reciprocity if
citations were made uniformly at random, but this doesn't necessarily imply
a tit-for-tat return of citations.  A citation is presumptively an
indication that two papers fall in similar subject areas, and thus the
presence of a citation greatly increases the chances that the authors are
working in the same area, which in turn increases the likelihood of
citation in general and therefore the chances of reciprocated citation.  In
the case of previous coauthors the chances of working in the same field are
likely even higher.  Unfortunately, we currently do not have any model of
the citation process detailed enough to make a quantitative prediction of
the size of this effect against which we could compare our measurements to
test for significance.

\subsection{Transitivity}

Transitivity, in the context of networks, refers to the observation that
``the friend of my friend is also my friend''~\cite{Wasserman94}.  In the
context of coauthorship, for example, it is observed that if A has
coauthored a paper with~B and B with~C, then A and C are more likely also
to have coauthored a paper.  One can define a so-called clustering
coefficient that quantifies this effect, measuring the average probability
that the friend of your friend is also your friend~\cite{Watts98}, and such
coefficients have been measured in many
networks~\cite{Banks96,Jin01,Davidsen02,Newman03}.  Typically one finds
that the values are significantly higher than one would expect if network
connections were made purely at random, and our coauthorship network is no
exception.  For the data set studied here we find a clustering coefficient
of $0.212$, which is comparable with other figures reported for
coauthorship networks~\cite{Newman01}.

In this case, however, the nature of the data set allows us to go further.
The conventional explanation for high transitivity in networks relies on a
triadic closure mechanism, under which two authors who share a common
coauthor are more likely to collaborate in future, perhaps because they
revolve in the same circles, attend the same conferences, work at the same
institution, or are introduced to one another by their common
acquaintance~\cite{Newman01a, Davidsen02, Holme02}.  The present data set's
time-resolved nature allows us to test this hypothesis directly.  We can
calculate what fraction of the time individuals who share a common coauthor
but have not previously collaborated themselves later write a paper
together.  When we make this measurement for the Physical Review data we
find the fraction of such author pairs to be only 0.0345---a much smaller
fraction than the clustering coefficient of the whole network reported
above.  One reason for this small figure is that a large fraction of the
transitivity seen in coauthorship networks comes from papers with three or
more authors, which automatically contribute closed triads of nodes to the
coauthorship network.  Such triads however are excluded from our
calculation of the probability of later collaboration.  The large
difference between the two probabilities we calculate implies that only a
small fraction of the network transitivity comes from true triadic closure
processes.

\begin{figure}
  \includegraphics[width=\columnwidth,clip=true]{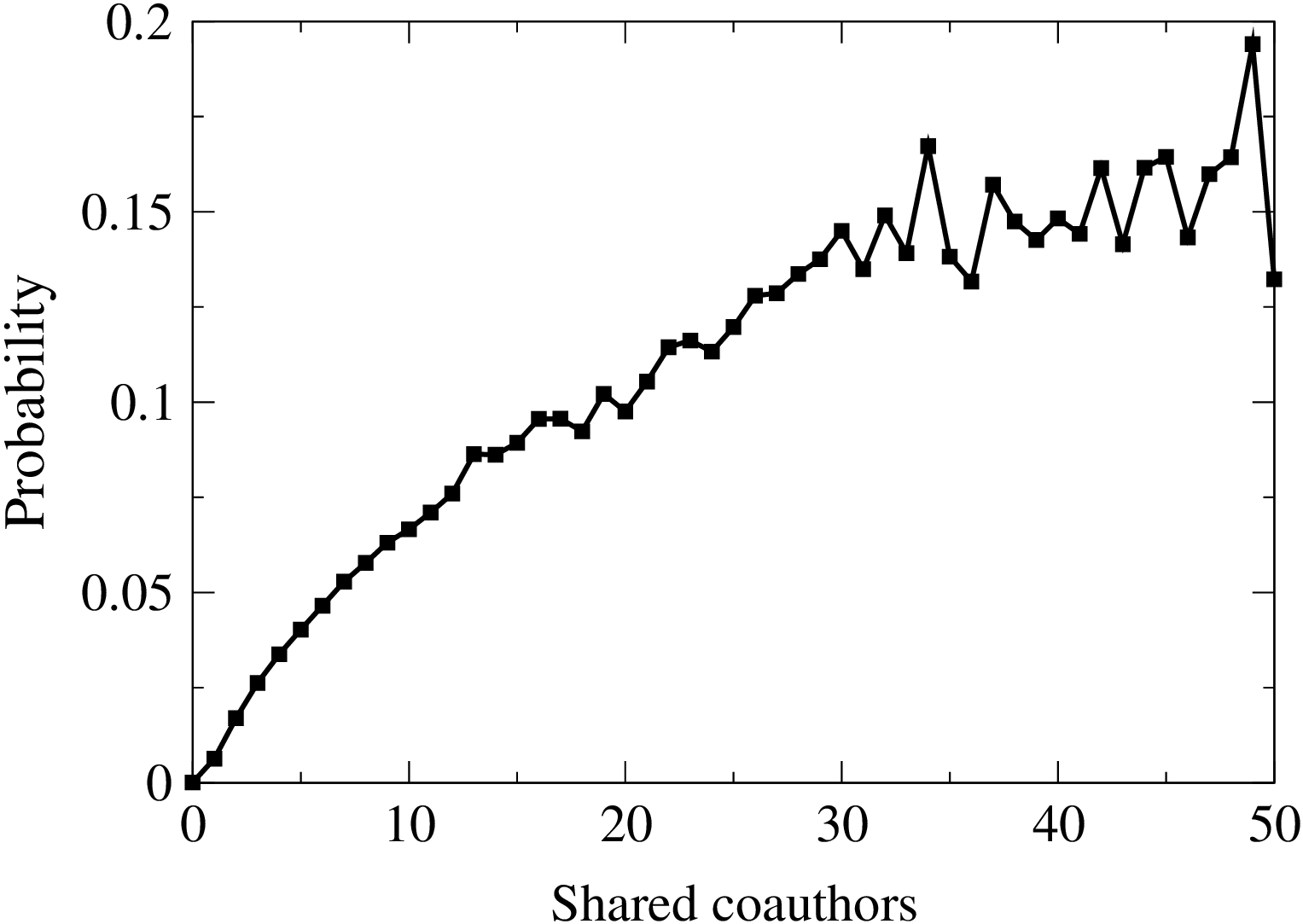}
  \caption{Probability of future coauthorship with another author as a
    function of the number of shared coauthors. The number of shared
    coauthors is counted at the time of first coauthorship or the date of either coauthor's last published paper, whichever comes first.}
  \label{fig:triad_prob}
\end{figure}

Nonetheless, the triadic closure process does appear to be present in our
data set.  Figure~\ref{fig:triad_prob} shows the probability of future
coauthorship between two individuals as a function of their number of
common coauthors, and we see that the probability increases sharply, a
finding that is consistent with previous results
\cite{Newman01a,Bloznelis12}.

\section{Conclusion}
In this paper we have analyzed a large data set from the Physical Review
family of journals, taking a network perspective.  Rather than focus solely
on either citation or coauthorship networks, as most previous studies have
done, we have instead combined the two, which allows us to study questions
about the ways in which people---and not just papers---cite one another,
and the extent to which scientists collaborate with those they cite or cite
those with whom they collaborate.  The time-span of the data set is
unusually large, covering more than a century of publication, which allows
us to study long-term changes in collaboration and citation patterns that
are not accessible with smaller data sets.

Our main findings are that the Physical Review appears to be growing
exponentially, with a doubling rate slightly less than 12 years, and the
number of citations per paper within the journals also appears to be
growing.  The fraction of self-citations and citations among coauthors is
more or less constant over time, and authors tend to cite their own papers
sooner after publication than do their coauthors, who in turn cite sooner
than non-coauthors.  We observe a strong tendency towards reciprocal
citations, researchers who cite another author often receiving a citation
in return later on, with especially high rates for citations between
coauthors.  Contrary to some previous claims~\cite{Newman01a, Davidsen02,
  Holme02}, however, there is only a small triadic closure effect in the
coauthorship patterns; two researchers who share a common coauthor but have
never collaborated themselves have only a rather small probability of
collaborating in future---about 3.5\%. This number is nonetheless much
higher than the probability for two randomly chosen researchers, and
moreover increases sharply as the number of common coauthors increases.

There are many other questions that could be addressed with this data set,
the unusually long time-span and combination of publication and citation
data opening up a variety of possibilities.  For instance, we know which
papers are published in which of the various Physical Review journals, and
hence we have a crude measure of paper topic, which would allow us to
answer questions about how the patterns of coauthorship and citation vary
between fields within physics.  We could also study geographical variations
by making use of the data on authors' institutional
affiliations~\cite{Pan12}.  Our analysis of long-term historical trends
could also be extended; for the researcher interested in the history of US
physics, there are, no doubt, many interesting signatures of historical
events hidden within the data.  The data set also offers the possibility of
tracking the careers of individual scientists, possibly over long periods
of time, or of tracking research on a particular topic.  And finally, any
of our analyses could be extended to data sets that cover other journals or
fields other than physics, if and when such data become available.  All of
these would make excellent subjects for future investigation.

\begin{acknowledgments}
This work was funded in part by the National Science Foundation under
grant DMS--1107796 (BB and MEJN) and an NSF IGERT Fellowship (TM).
\end{acknowledgments}

\appendix
\section{Data processing}
\label{app:data_processing}
As mentioned in the main text, we performed some pre-processing on the raw
Physical Review data to disambiguate author names and remove extreme
outliers.  This appendix describes the steps taken.

\subsection{Author name disambiguation}
The data were supplied in two blocks: (1)~a list of papers with associated
information, such as authors, author affiliation, journal, and year of
publication; (2)~a list of citations, using unique paper identifiers that
correspond to entries in the first block.  There are, however, no unique
identifiers for authors that are consistent between papers, making
unambiguous author identification difficult.  Not all authors use the same
form for their name on every publication, and there are many examples of
distinct researchers with the same name.  Before using the data set,
therefore, we made an effort to associate names of authors with unique
people.  As in previous work on author disambiguation, our process starts
by assuming every name on every paper to represent a different individual,
then computes a number of measures of author similarity and assumes authors
who are sufficiently similar by these measures to be the same person.
After completing this disambiguation process we checked a subset of the
results by hand to estimate error rates for the process and found that it
performs well.  Details are as follows.

Our approach relies not only on the author names themselves to establish
similarity, but also on collaboration patterns and institutional
affiliation, since authors with similar names who have many of the same
collaborators or who are at the same institution are more likely to be the
same person.  Affiliation information, however, like the author names
themselves, tends to be ambiguous and inconsistent, so our first step is to
combine affiliations that are deemed similar enough.  We measure similarity
using a variant of edit distance applied to the affiliation text strings,
implemented using the Python difflib library.

Once the affiliations are processed in this way, we process the author
names as follows:
\begin{enumerate}
\item We combine all authors with identical names who share an
  institutional affiliation.  It appears to be uncommon for two physicists
  at the same university to publish under identical names, so this
  seems to be a safe step.
\item We find author pairs with similar but not identical names.  Our
  criterion for similarity at this stage is that authors should have
  identical last names and compatible first/middle names (i.e.,~identical
  if fully written out, or compatible initials where initials are used).
  Also authors should not have published together on the same paper (which
  rules out, for example, family members with similar names who publish
  together).  For all pairs with similar names we then calculate a further
  similarity measure based on how many affiliations they share, how many
  coauthors they share, whether their full names are identical, and whether
  they have published in the same journal.  Authors with a high enough
  similarity are combined, most similar pairs first.
\end{enumerate}

We have tested the accuracy of this process by drawing two lists at random
from its output, the first containing 79 instances in which authors with
similar names have been combined into a single author, and the second
containing 111 instances in which they have not.  We then performed, by
hand, a blind search---without knowing the choice the algorithm has
made---for publicly available on-line information about the names in
question, to determine whether they do indeed represent the same or
distinct researchers.  We find the false positive rate to be 3\% (i.e.,~3\%
of pairs are incorrectly judged to be the same person when in reality they
are distinct) and the false negative rate to be 12\%.

We also tested the effect on our results of the disambiguation process by
calculating a number of the statistics reported in this paper both for the
disambiguated data and for the raw data set before disambiguation, in which
we naively assume that every unique author string represents a unique
author and every pair of authors with the same string are the same person.
We found substantial differences between the two in some of the most basic
statistics, such as total number of distinct authors: the number was
$328\,938$ in the raw data set, but fell to $235\,533$ after
disambiguation.  On the other hand some other statistics changed very
little, indicating that these are not particularly sensitive to details of
author identification.  For example, the clustering coefficient changes
from $0.222$ in the raw data set to $0.212$ in the disambiguated data set.

\begin{figure}[t]
  \includegraphics[width=\columnwidth,clip=true]{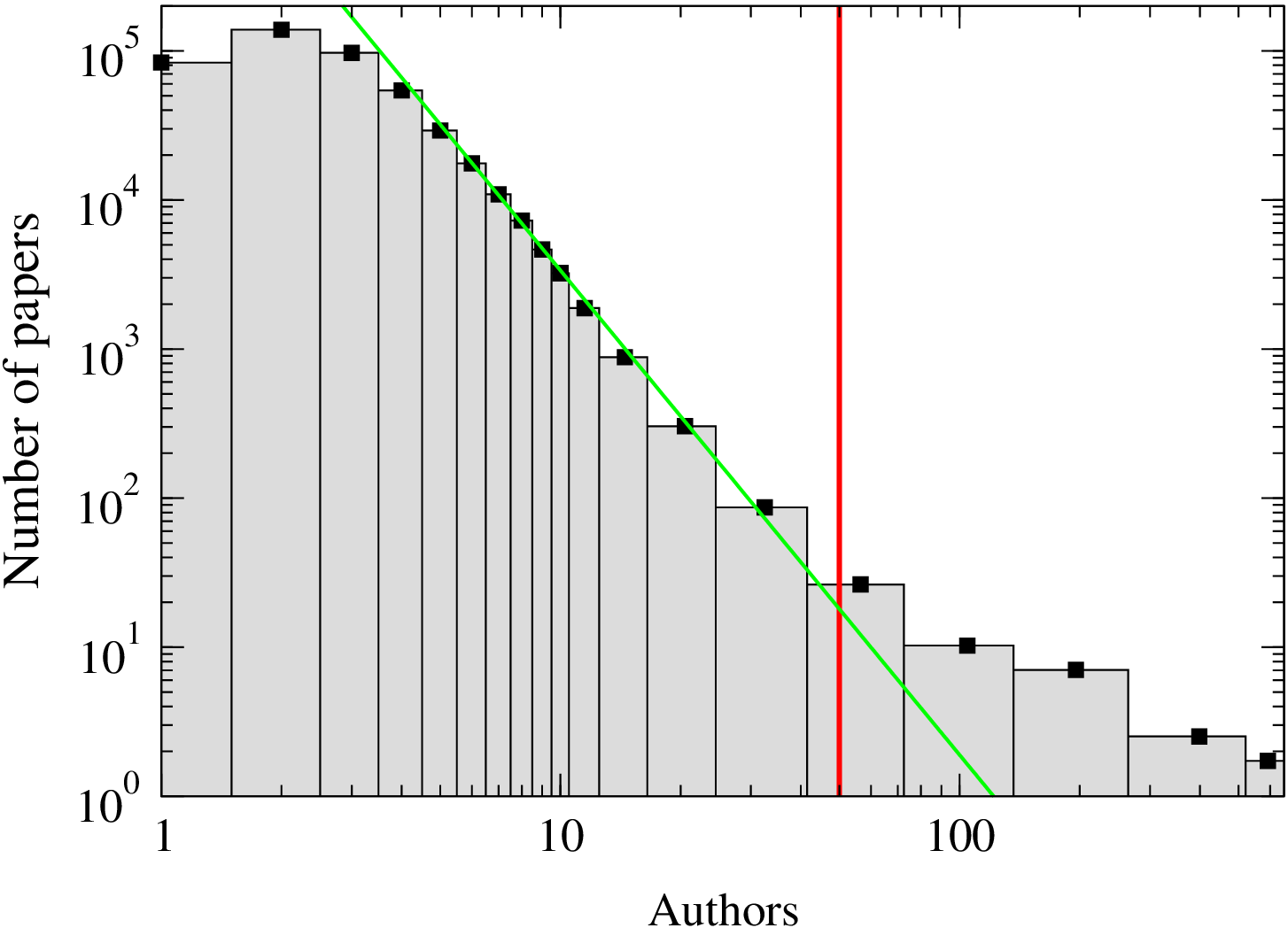}
  \caption{Histogram of the number of papers with a given number of
    authors.  The vertical line falls at fifty authors and corresponds
    roughly to the point at which the distribution deviates from the
    power-law form indicated by the fit.  The data for ten authors and more
    have been binned logarithmically to minimize statistical fluctuations.}
  \label{fig:papers_auths}
\end{figure}

\subsection{Data culling}
In addition to author disambiguation we cull the data according to a few
simple rules.  There are a number of papers in the data set that have no
authors listed, primarily editorials and other logistical articles without
scientific content.  These we remove entirely.  As mentioned in the text,
we also identify all papers with fifty or more coauthors, and many of our
calculations are performed in two versions, with and without these
papers.  The choice of fifty authors as the
cutoff point was made by inspection of the distribution of author numbers
shown in Fig.~\ref{fig:papers_auths}.  As the figure shows, the number of
papers with a specific number of coauthors appears, roughly speaking, to
follow a power law (in agreement with some previous
studies~\cite{Newman01b}, but not others~\cite{Hsu09}), but there is a
marked deviation from the power-law form for the highest numbers of
coauthors, above about fifty, indicating potentially different statistical
laws in this regime, and possibly different underlying collaborative
processes.

We also removed from the data a small number of citations.  In a few cases
a paper is listed as citing itself, which we assume to be an error.  In a
number of other cases papers cite others that were published at a later
time, which violates causality.  These too are assumed to be erroneous and
are removed.  Finally, the data indicate that some papers cited the same
other paper several times within the one bibliography; such multiple
citations we count as a single citation.

\end{document}